# Improvement of Morphology and Free Carrier Mobility through Argon-Assisted Growth of Epitaxial Graphene on Silicon Carbide

J.L. Tedesco, B.L. VanMil, R.L. Myers-Ward, J.C. Culbertson, G.G. Jernigan, P.M. Campbell, J.M. McCrate, S.A. Kitt, C.R. Eddy, Jr., and D.K. Gaskill

U.S. Naval Research Laboratory, 4555 Overlook Avenue SW, Washington, DC 20375

Graphene was epitaxially grown on both the C- and Si-faces of 4H- and 6H-SiC(0001) under an argon atmosphere and under high vacuum conditions. Following growth, samples were imaged with Nomarski interference contrast and atomic force microscopies and it was found that growth under argon led to improved morphologies on the C-face films but the Si-face films were not significantly affected. Free carrier transport studies were conducted through Hall effect measurements, and carrier mobilities were found to increase and sheet carrier densities were found to decrease for those films grown under argon as compared to high vacuum conditions. The improved mobilities and concurrent decreases in sheet carrier densities suggest a decrease in scattering in the films grown under argon.

## Introduction

Graphene films have been identified as having a wide range of unique physical (1-4), chemical (1,5-6), and electronic properties (4,7-12) that make graphene highly attractive for use in a wide range of electronic and sensor applications (4-6,13-14). Furthermore, 300 K mobilities greater than 10,000 $cm^2V^{-1}s^{-1}$ have been reported (7,15). These mobilities are significantly greater than those of other elemental semiconductors (16) and comparable to or greater than mobilities for compound semiconductors (17). Graphene can also be formed epitaxially over large areas by graphitizing silicon carbide (SiC) in a chemical vapor deposition (CVD) reactor (15,18). The high mobility and growth over large areas make epitaxial graphene viable for a wide range of electronic applications.

It is well known that the ideal surface for conventional semiconductor processing is one that is morphologically uniform and smooth. Previous studies have shown that the underlying morphology of the SiC is rough following the CVD graphitization process (14-15,18-19). Step bunching occurs on both faces and, in the case of C-face growth, steps become erratic in shape. Furthermore, the topographic features of the C-face epitaxial graphene films can lead to height differences of up to 100 nm. It has been shown that silicon begins to sublime from SiC at a higher temperature in an inert environment than in vacuum (20-21). This suggests that forming graphene in an inert ambient, such as argon, may result in the control of substrate morphology, and therefore, the graphene morphology. In fact, growth under various partial pressures of argon has been shown to lead to improved morphology of the graphene films (22-23). In this study, the morphological and electrical properties of epitaxial graphene films grown under argon will be compared directly to the properties of epitaxial graphene films grown under high vacuum in the same CVD reactor.

## Experimental Procedure

Semi-insulating, on-axis (0°±0.5°), 76.2 mm diameter 4H- and 6H-SiC(0001) wafers, epi-ready with chemical-mechanical polished (CMP) surfaces, were obtained from Cree (4H) and II-VI, Inc. (6H) and subsequently diced into 16 × 16 mm$^2$ samples. Both the C- and Si-faces were used for growth. A previous study determined that the substrate polytype had no systematic effect on either the morphological or the electrical properties (15), thus, while growth runs contained both C- and Si-face samples, they did not always contain both 4H and 6H samples. The samples were chemically cleaned *ex situ* (24) prior to loading into the Aixtron/Epigress VP508 Hot-Wall CVD reactor. Once the samples were loaded and adequate chamber vacuum was achieved (<4.0 × 10$^{-7}$ mbar), processing was started by hydrogen etching the sample surface. The hydrogen etch step was performed for 5 to 20 minutes at a pressure of 100 mbar and temperature of 1600°C in order to remove the CMP polishing damage from the surface. It has been estimated that the hydrogen etching under these conditions removes in excess of 300 nm of the original top surface, which is generally sufficient to remove surface polishing damage and leave behind a well-ordered, uniformly-stepped surface (15,24).

For growth runs under high vacuum following the hydrogen etch step, the temperature was maintained at 1600°C and argon flowed into the chamber at 100 mbar to flush out the hydrogen. The growth chamber was pumped using the reactor process pump (Ebara A25S). The argon flush was discontinued after 10 minutes and the chamber was evacuated using a turbopump (Pfeiffer TMH 521) as the substrate temperature was ramped to the growth target. The pressure during vacuum synthesis began in the 10$^{-4}$ mbar range and steadily decreased over the length of the run, generally ending in the low-10$^{-5}$ to mid-10$^{-6}$ mbar range.

For argon growth runs, the 100 mbar argon pressure was maintained after the hydrogen etch step while the temperature was ramped to the growth target. At the end of the argon growth runs the substrate temperature was reduced to below 1500°C while the argon was evacuated. The growth temperature for both vacuum and argon growth runs ranged from 1500 to 1600°C for the 60 to 120 minute growth runs. And for both growth cases, samples were cooled under turbopumped vacuum for several hours (typically overnight).

Following growth, samples were removed from the reactor and graphene was confirmed by a finite electrical resistance (range: ~10$^2$ to 10$^5$ Ω) of the epitaxial surface. Raman spectroscopy was accomplished using a 150 mW 532 nm laser, 0.5 m single-pass spectrometer (Acton SP2500) and nitrogen-cooled CCD array (Princeton Instruments Spec-10) and spectra of the vacuum- and argon-grown conductive films confirmed graphene by the 2D peak at ~2700 cm$^{-1}$ (25). The surface morphologies of the graphene films were characterized by Nomarski interference contrast microscopy and tapping mode atomic force microscopy (AFM: Digital Instruments Dimension 3100). The Hall effect mobilities were measured at both 300 K and 77 K using a van der Pauw configuration with copper pressure clips serving as contacts at the corners of the films. Measurement currents used ranged from 1 to 100 μA and the magnetic field was 2,060 G.

## Results

The effect of argon growth on the morphologies of the epitaxial graphene and the underlying SiC substrate depended on whether the graphene film was grown on the C-face or the Si-face of SiC. The following discusses and compares the results of C-face growths in argon and vacuum. After heating to 1500°C for 60 minutes under argon, the sample's morphology did not appear significantly different from the morphology of a C-face SiC sample following hydrogen etching, as shown in Figs. 1(a) and 1(b), respectively. Ohmmeter tests indicated that the surface was non-conductive, and Raman spectroscopy measurements showed no evidence of the graphene 2D peak, demonstrating that graphene had not formed. After heating to 1550°C for 60 minutes under argon, ohmmeter tests on the sample also measured an infinite resistance. However, surface morphology by Nomarski microscopy and AFM, was consistent with epitaxial graphene as incomplete patches across the surface, as shown in Figs. 2(a) and 2(b), respectively. Graphene identification was confirmed by Raman spectroscopy, as shown in Fig. 2(c) which compares the 2D peaks from a graphene patch and nearby bare SiC region, such as those shown in Fig. 2(a) labeled "EG" and "SiC". Epitaxial graphene can also be distinguished from the SiC by the presence of randomly-oriented ridges 25 to 100 nm in height, depicted as bright white lines in the AFM image of the graphene film in Fig. 2(b); these have been observed previously on C-face growth in vacuum ("giraffe stripes") (15). Furthermore, the epitaxial graphene film grown under argon at 1550°C was significantly different from an epitaxial graphene film grown under vacuum at the same temperature; the latter is shown in Figs. 2(d) and 2(e). The epitaxial graphene film grown under vacuum was continuous and was marked by a high density of pits formed in the underlying SiC during graphene growth. These pits appear as irregular black features in Figs. 2(d) and 2(e). Following heating to 1600°C for 60 minutes under argon, the resistance of the epitaxial graphene film was finite and Raman spectra displayed the 2D peak, indicating the presence of graphene. Nomarski and atomic force microscopies showed that the film was continuous; examples are in Figs. 3(a) and 3(b). In contrast, the epitaxial graphene film grown at 1600°C under argon was less rough than an epitaxial graphene film grown in vacuum at the same temperature, shown in Figs. 3(c) and 3(d). The rms roughness of the argon-grown epitaxial graphene film is 10 nm, while the rms roughness of the vacuum-grown epitaxial graphene film is 20 nm. The difference in rms roughness is due to the improved morphology of the underlying SiC substrate during argon growth; in particular, the argon-grown sample does not contain a high density of irregular pits in the underlying SiC.

The growth of the epitaxial graphene is different on the Si-face than on the C-face and the following discusses and compares Si-face growths in vacuum and argon. Following heating at 1500°C for 60 minutes, the resistance of the sample surface was infinite and Raman spectroscopy indicated that no graphene was present on the surface; however, the morphology of the surface did change due to the heating. This is shown clearly by comparing Fig. 4(a), the 1500°C argon morphology (after the hydrogen etch) with Fig. 4(b), the hydrogen etched morphology. Figure 4(a) shows step bunches with heights of 6 to 9 nm, which are absent in the hydrogen etched surface (~1 nm step heights) shown in Fig. 4(b). This is likely the onset of step bunching that is observed after argon growths at higher temperatures and vacuum growths at 1500°C. The resistances of the films grown at 1550° and 1600°C were both finite, and Raman spectroscopy measurements indicated that the films were graphene. No significant differences in the morphology of the epitaxial graphene films grown under argon or vacuum were found. Evidence of this is shown in Fig. 5 comparing the Nomarski and atomic force microscopy images samples grown under argon and under vacuum at 1600°C.

Growth under argon significantly increased the film mobility for all but one of the C-face and all but two of the Si-face samples studied, as compared to high vacuum films grown at the same temperature and for the same length of time. Furthermore, growth under argon also decreased the sheet carrier densities for both C- and Si-face samples as compared to high vacuum C- and Si-face samples that otherwise had the same growth conditions. The maximum mobilities and minimum sheet carrier densities found for the C- and Si-face graphene films grown under the same growth temperatures and times are shown in Table I. The best C-face graphene samples grown under argon showed 300 K mobilities that were several times greater than the mobilities of those films grown in vacuum, with an order of magnitude decrease in 300 K sheet densities. The behavior of the mobilities and sheet densities for the best Si-face films was similar to the behavior observed for the C-face films. The Si-face epitaxial graphene films grown under argon showed increases in 300 K mobilities that were 2 to 3 times greater than the mobilities of the films grown under vacuum and the 300 K sheet densities decreased by over an order of magnitude. Samples generally behaved in a similar manner at 77 K as they did at 300 K. The best C-face samples grown under argon exhibited increases in 77 K mobility of several times over the mobilities of vacuum-grown samples, while the sheet carrier densities decreased by approximately an order of magnitude. The best argon-grown Si-face samples showed 77 K mobility increases of 2 to 3 times over the mobilities of vacuum-grown films. Again, the decrease in sheet density for the argon-grown Si-face films was approximately an order of magnitude relative to the sheet densities of the vacuum-grown Si-face films.

**TABLE I.** The maximum mobilities and the minimum sheet carrier densities measured for $16 \times 16$ mm$^2$ epitaxial graphene films grown under the growth conditions shown.

| Graphene Film Type | Growth conditions | | | 300 K | | 77 K | |
|---|---|---|---|---|---|---|---|
| | Temperature (°C) | Time (min) | $P_{argon}$ (mbar) | Maximum mobility (cm$^2$V$^{-1}$s$^{-1}$) | Minimum sheet carrier density (cm$^{-2}$) | Maximum mobility (cm$^2$V$^{-1}$s$^{-1}$) | Minimum sheet carrier density (cm$^{-2}$) |
| C-face | 1500 | 60 | 100 | 0 | n/a | 0 | n/a |
| | | | 0 | 2,366 | $3.6 \times 10^{13}$ | 5,252 | $1.6 \times 10^{13}$ |
| | 1600 | 60 | 100 | 2,837 | $2.0 \times 10^{13}$ | 5,919 | $9.6 \times 10^{12}$ |
| | | | 0 | 244 | $4.7 \times 10^{14}$ | 636 | $1.3 \times 10^{14}$ |
| | 1600 | 120 | 100 | 3,168 | $1.9 \times 10^{13}$ | 7,197 | $1.0 \times 10^{13}$ |
| | | | 0 | 949 | $1.5 \times 10^{14}$ | 1,240 | $8.3 \times 10^{13}$ |
| Si-face | 1500 | 60 | 100 | 0 | n/a | 0 | n/a |
| | | | 0 | 137 | $1.1 \times 10^{13}$ | 385 | $3.5 \times 10^{12}$ |
| | 1600 | 60 | 100 | 453 | $4.0 \times 10^{12}$ | 838 | $2.0 \times 10^{12}$ |
| | | | 0 | 187 | $8.2 \times 10^{13}$ | 794 | $1.6 \times 10^{13}$ |
| | 1600 | 120 | 100 | 627 | $3.3 \times 10^{13}$ | 2,647 | $1.0 \times 10^{12}$ |
| | | | 0 | 380 | $7.5 \times 10^{13}$ | 1,363 | $2.1 \times 10^{13}$ |

There appears to be a general trend between increasing mobility and decreasing carrier density for all films independent of growth ambient. The mobilities and corresponding sheet carrier densities for all samples measured at 300 K have been plotted in Fig. 6. The dashed lines are linear fits to this data and correspond to the growth face of the underlying SiC, regardless of the growth ambient. The trend is more pronounced for epitaxial graphene films grown on the C-face than the trend for graphene films grown on the Si-face. However, for both C- and Si-face samples, the relationship between mobility and sheet density for epitaxial graphene films is similar for films grown under argon and vacuum. The similar relationships suggest that growth under argon is not changing either the growth mode or the electronic structure of the graphene film, as will be discussed in the next section.

## Discussion

It has been reported that graphene formation on SiC begins at ~800°C to 1150°C in ultrahigh vacuum conditions ($10^{-9}$ to $10^{-10}$ mbar) (23,26-27). However, it has also been recently reported that under argon pressures up to 900 mbar, the sublimation of silicon is suppressed until ~1500°C (23). This suppression of silicon sublimation is consistent with the images shown in Figs. 1 and 4, which show that epitaxial graphene did not grow on either face of SiC at 1500°C. Therefore, the conditions previously reported for optimal graphene growth on the C-face under high vacuum conditions (15) may not be valid for growth under argon. It was previously reported that longer duration growths at 1600°C were required for optimal graphene growth under vacuum on the Si-face (15); therefore the higher sublimation temperature in argon would most likely not affect optimal epitaxial graphene growth on the Si-face as much as it does on the C-face.

The rate of epitaxial graphene growth is suppressed by the argon ambient at 1550°C. This can be seen by comparing the topography of the C-face graphene films grown in argon in Figs. 2(a) and 2(b) with the C-face graphene films grown in vacuum in Figs. 2(c) and 2(d). However, as shown in Fig. 2(b), ridges that were observed during vacuum growth at 1550°C (15,19) still appeared on the C-face films during growth under argon. Ridges were also present in the epitaxial film grown at 1600°C under argon, as shown in Fig. 3(b). Furthermore, the density of ridges for films grown under argon does not appear to be lower than the density of ridges under vacuum at both 1550°C and 1600°C. Thus, the presence of equivalent densities of ridges suggests that their formation is not related to the film growth rate at either 1550°C or 1600°C. In contrast, on the Si-face, epitaxial graphene growth morphologies are not significantly affected by growth under argon consistent with previous observations that growth is already slow and self-terminating under vacuum (11). The similarities in the graphene growths in vacuum and argon suggest that the growth mechanism for epitaxial graphene has only been slowed, but not fundamentally altered, by the argon process.

For epitaxial films on both the C- and Si-faces, both electron and hole mobilities were measured for both argon-grown and vacuum-grown films. However, at present, the differences between them have been neglected and the data has been plotted as if the dominant carriers in each sample were identical. Such a consideration is reasonable given that graphene is ambipolar and has a linear dispersion relation (10), implying that there should be no intrinsic distinction between electrons and holes in epitaxial graphene other than electrical charge.

The data shown in Fig. 6 indicates that 300 K growth under argon appears to increase the mobility for most samples, and decrease the sheet carrier density for all samples. Furthermore, the behavior of mobility as a function of sheet carrier density measured from the argon-grown samples is consistent with the behavior observed for the vacuum-grown samples, also shown in Fig. 6. The fact that the data for argon-grown samples show similar trends to the vacuum-grown samples suggests that the scattering mechanisms affecting mobility in the epitaxial graphene films are not significantly changed by growth under argon.

The reduction in sheet carrier densities of the argon-grown samples may be due to the removal of impurities consistent with previous studies suggesting that annealing graphene in argon removed volatile impurities and contaminants from exfoliated graphene films (28-29). The reduction in sheet density implies that there would be a reduction in carrier-carrier scattering and this could result in increased mobility. In any case, the source of free carriers in epitaxial graphene films is currently an open question and investigations are currently in progress.

## Conclusions

This work shows that epitaxial graphene can be grown on SiC at temperatures greater than 1500°C in an argon ambient of 100 mbar. The morphologies of graphene grown on the C- and Si-faces under argon or high vacuum are significantly different; the former is dominated by ridges. While the morphologies of the Si-face films do not change significantly under argon, the morphologies of the C-face films are significantly improved through the elimination of surface pitting. Furthermore, the rate of graphene formation on the C-face at 1550°C is significantly reduced under argon as compared to growth under vacuum. These results point to a growth mechanism in argon that is similar to the vacuum-based approach.

Graphene growth under argon instead of high vacuum at 1600°C led to increased mobility and decreased sheet carrier density. The increased mobility suggests that the growth under argon may be removing impurities. Furthermore, the relationships between mobility as a function of sheet carrier density measured from films grown under argon and under high vacuum are similar. This similarity suggests that the scattering mechanisms affecting the mobility of argon-grown graphene films are not significantly different than those scattering mechanisms for vacuum-grown films.

## Acknowledgements

The authors acknowledge support from the Office of Naval Research. JLT and BLV also acknowledge support from the American Society for Engineering Education—Naval Research Laboratory Postdoctoral Fellowship Program. JMM and SAK also acknowledge the support of the Naval Research Enterprise Intern Program conducted by the American Society for Engineering Education at the Naval Research Laboratory.

**Figures**

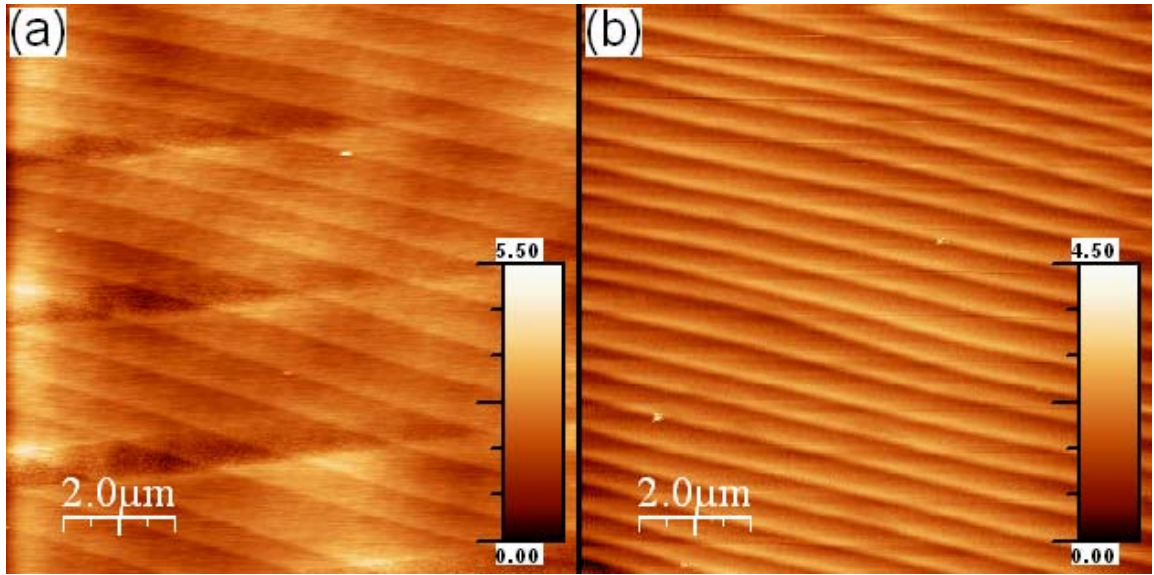

Figure 1. Atomic force microscopy images of the topographies of (a) a C-face surface following heating under argon at 1500°C and (b) a C-face SiC surface following hydrogen etching. The height scales are in nanometers.

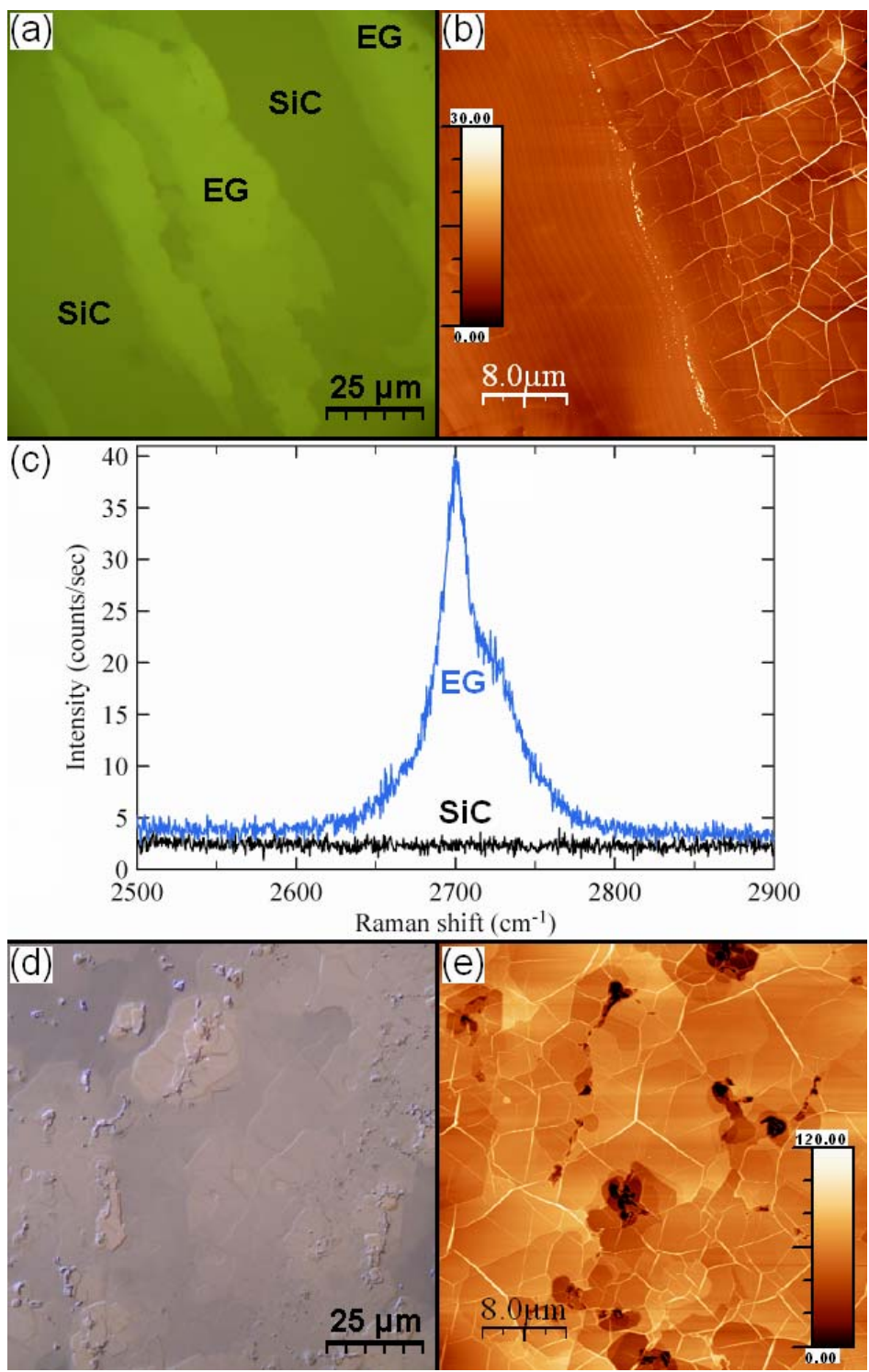

Figure 2. (a) Nomarski micrograph and (b) AFM image of the surface of a C-face graphene film grown at 1550°C under argon. Patches of epitaxial graphene in (a) are labeled "EG" and bare SiC is labeled "SiC." (c) Comparison of Raman spectra from a region containing patches of epitaxial graphene. The spectrum labeled "EG" corresponds to the spectra on a graphene patch, while the spectrum labeled "SiC" corresponds to the bare SiC next to the patch. (d) Nomarski micrograph and (e) AFM image of the surface of a C-face graphene film grown at 1550°C in vacuum. The irregular black shapes in (e) are pits in the SiC underlying the graphene film. The height scales are in nanometers.

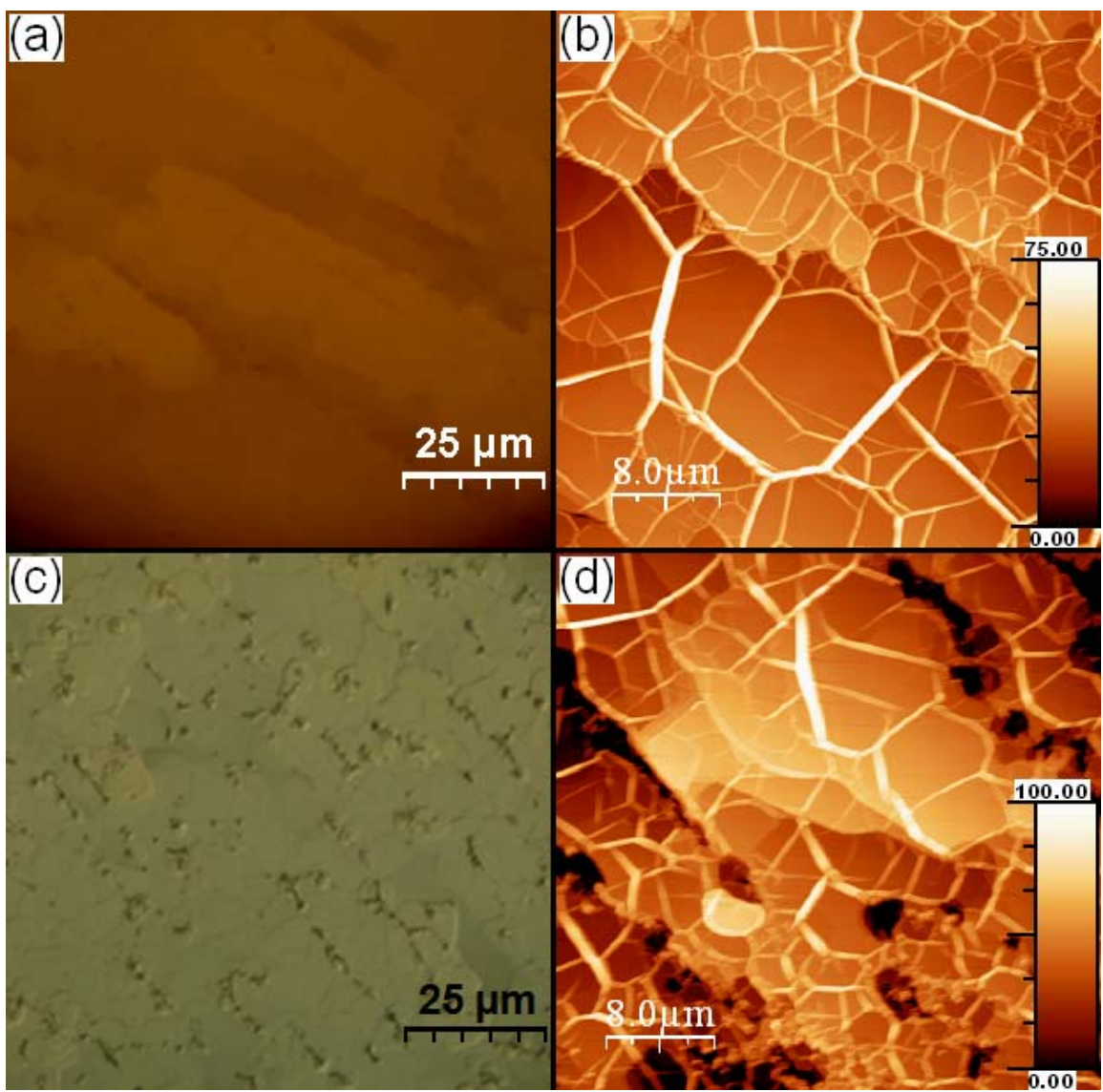

Figure 3. (a) Nomarski micrograph of the surface of a C-face graphene film grown at 1600°C under argon. (b) Atomic force microscopy image of the topography of the same C-face film. (c) Nomarski micrograph of the surface of a C-face graphene film grown at 1600°C in vacuum. (d) Atomic force microscopy image of the topography of the same vacuum-grown C-face film. The height scales are in nanometers.

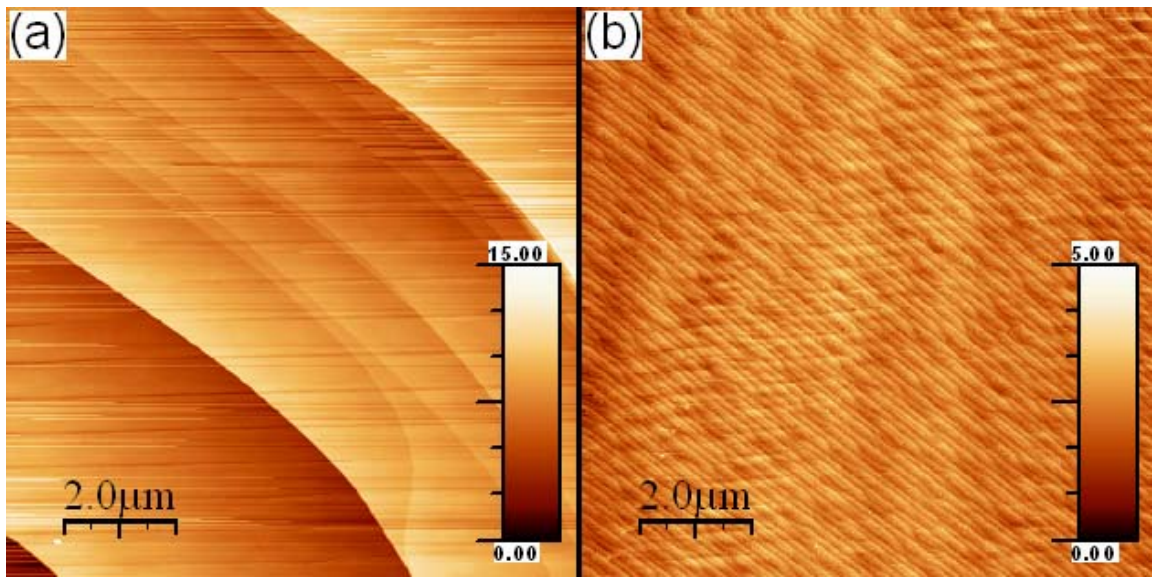

Figure 4. Atomic force microscopy images of the topographies of (a) a Si-face surface following heating under argon at 1500°C and (b) a Si-face SiC surface following hydrogen etching. The height scales are in nanometers.

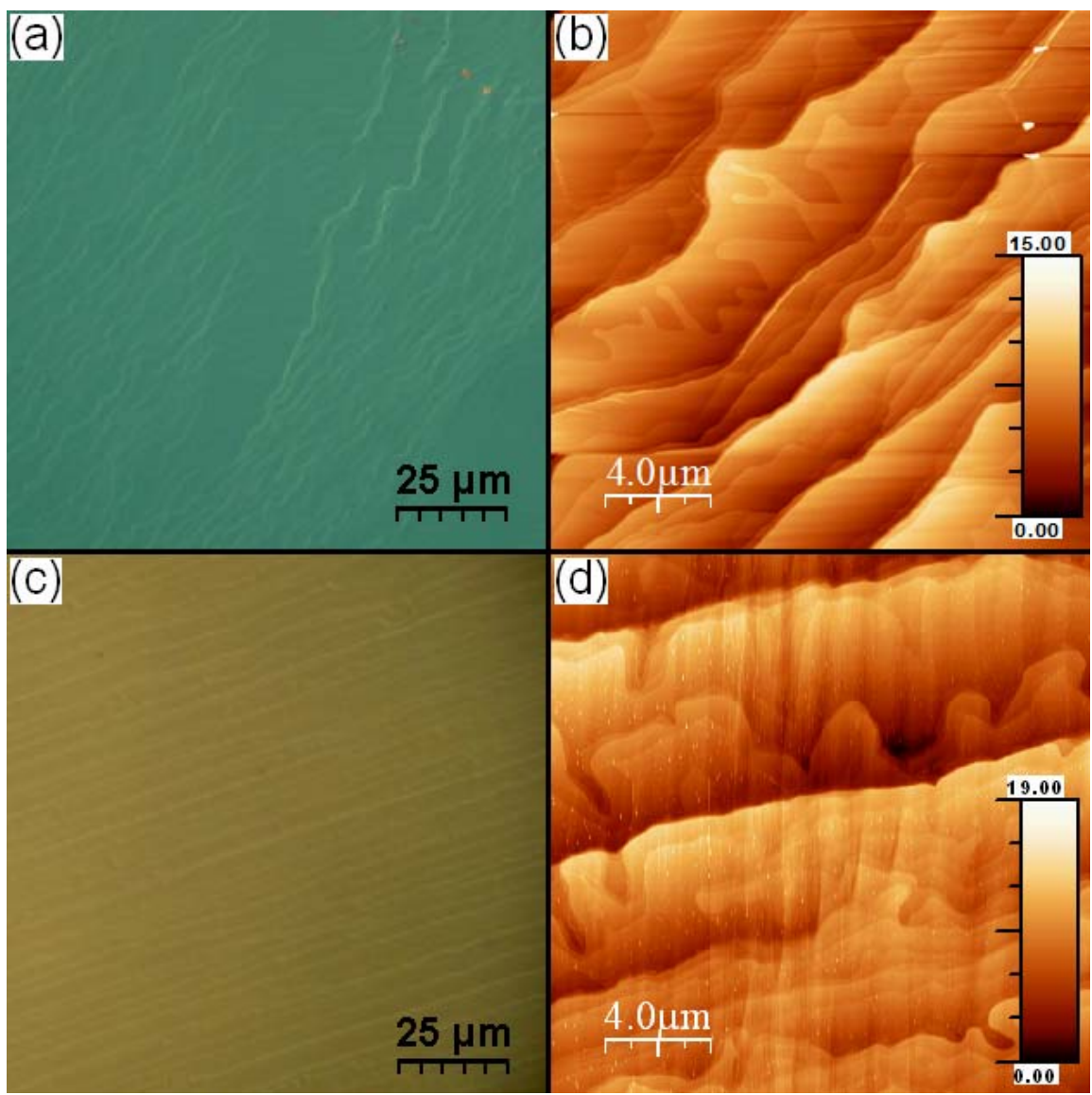

Figure 5. (a) Nomarski micrograph of the surface of a Si-face graphene film grown at 1600°C under argon. (b) Atomic force microscopy image of the topography of the same Si-face film. (c) Nomarski micrograph of the surface of a Si-face graphene film grown at 1600°C in vacuum. (d) Atomic force microscopy image of the topography of the same vacuum-grown Si-face film. The height scales are in nanometers.

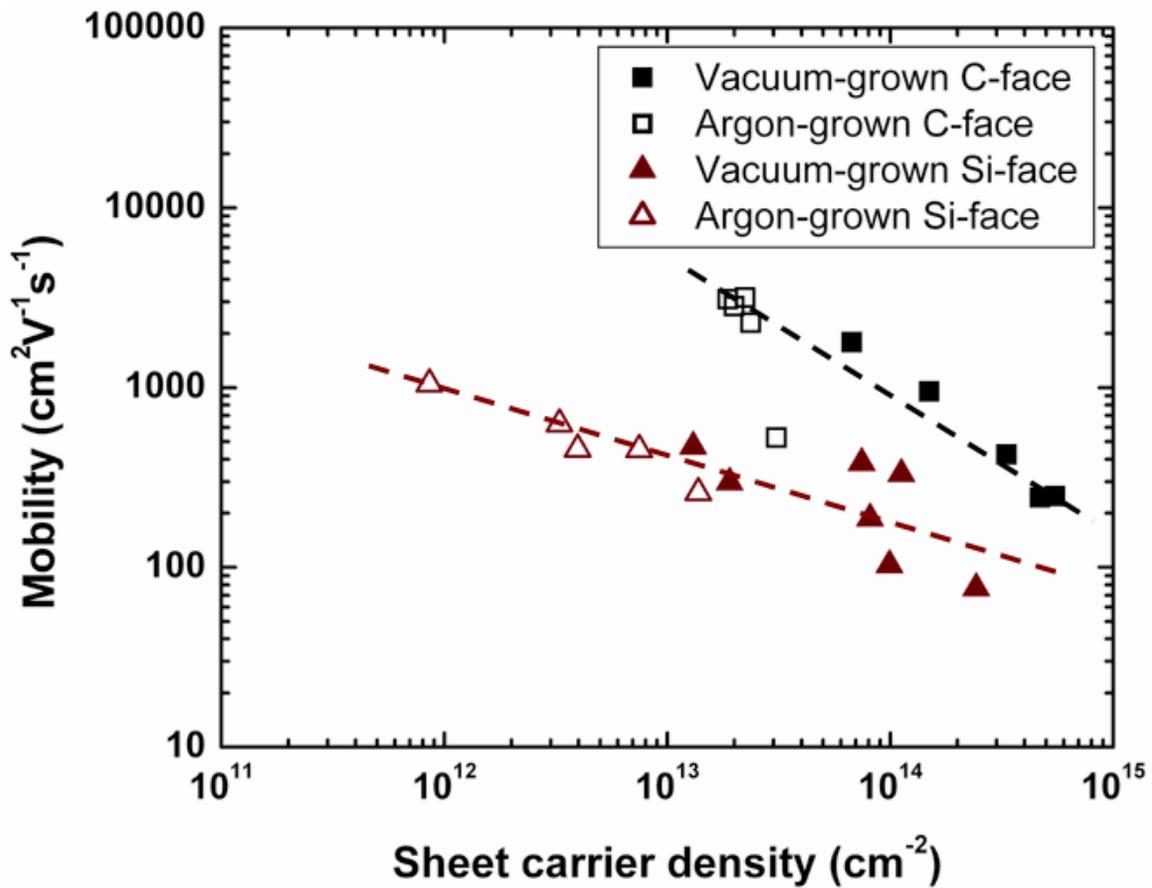

Figure 6. Hall effect mobilities and sheet carrier densities measured at 300 K for 16 × 16 mm$^2$ epitaxial graphene films grown on the C-face and Si-face grown under vacuum and under argon. The dashed lines are linear fits to the data for C- and Si-face samples.